\documentclass[preprint,prc,aps,tightenlines,floatfix,showpacs]{revtex4-1}
\usepackage{graphics}
\usepackage{bm}
\usepackage{amsmath}
\usepackage{amssymb}

\begin{document}
\title{Spin observables in $71A$ MeV $^{6,8}$He-hydrogen scattering}

\author{S. Karataglidis$^{(1,2)}$}
\email{stevenka@uj.ac.za}

\author{K. Amos$^{(2,1)}$}
\email{amos@physics.unimelb.edu.au}

\affiliation{$^{(1)}$
Department of Physics, University of Johannesburg,
    P.O. Box 524, Auckland Park, 2006, South Africa}

\affiliation{$^{(2)}$
School of Physics,The University of Melbourne,
Victoria 3010, Australia}

\begin{abstract}
Predictions of  cross sections and analyzing  powers using $g$-folding
optical potentials  for the scattering of  $71A$ MeV $^{6,8}$He ions
from (polarized) hydrogen are compared with data.  A $g$-folding model
in  which  exchange  amplitudes are evaluated explicitly was used with
wave functions of  $^{6,8}$He  specified  from a no-core shell model
that used a complete $(0+2+4)\hbar\omega$ basis.  The analyzing powers reveal
some sensitivities to the details of the wave functions, especially in the
case of halo nuclei.
\end{abstract}
\maketitle

\section{Introduction}

It has been  possible, for quite some time, to  make predictions  of  nucleon-nucleus ($NA$),
elastic scattering observables~\cite{Am00}, and, under a distorted wave 
approximation (DWA), predictions also of observables from inelastic  scattering and charge
exchange reactions. Such analyses have been possible
using a completely microscopic formulation of  scattering~\cite{Am00}. 
As all details required in the calculations are preset, with no adjustable parameters, 
the results of these calculations are predictions of the relevant scattering observables.
Only one run of the scattering code (DWBA98 \cite{Ra98} in our case) is needed.
Complete  details as
well as  many examples of  use of this (coordinate  space) microscopic
model approach are to be  found in the review~\cite{Am00}.  Use of the
complex, nonlocal, nucleon-nucleus optical  potentials defined in that 
way    predicted cross sections and spin observables in good agreement  
with data from many stable nuclei ($^3$He to $^{238}$U) and for a wide 
range of energies (25 to 300~MeV).  Crucial to that success was use of
effective   nucleon-nucleon  ($NN$)  interactions  built   upon   $NN$  
$g$-matrices that are solutions of   Brueckner-Bethe-Goldstone   (BBG) 
equations for realistic starting free $NN$ interactions. The effective 
$NN$ interactions are complex, energy and density dependent, have central, 
two-nucleon tensor and  two-nucleon spin-orbit character,    with each 
component represented by a form factor formed as a mixture of Yukawa functions. Details 
of this Melbourne force are also given in the review~\cite{Am00}.  The $NA$ 
optical  potentials result from folding those effective  interactions 
with  the  one-body  (ground) state density matrices (OBDME)  of   the  
target nucleus, as obtained from a suitable model of structure with nucleonic
degrees of freedom.

Antisymmetrization  of the  projectile with  all target nucleons leads 
to exchange (knock-out) amplitudes, so making the microscopic  optical 
potential nonlocal. An important facet is that nonlocality should be 
treated explicitly in the  calculations of  scattering.   For brevity,  
the optical  potentials that result are called $g$-folding potentials.

The level of agreement found between predictions of scattering observables
made using the $g$-folding potentials has been excellent for stable nuclei
across the mass range. But, it is also noteworthy that the model has been applied to the
scattering of exotic nuclei from hydrogen \cite{Ka97,Ka00,La01,St02} with equal
success. Another application has been the prediction of integral observables
of elastic scattering, for both protons and neutrons, with quite good agreement
\cite{De01}. Thus, the method is known to give good predictions of angular-dependent
and integral observables, which is not guaranteed when the usual phenomenological
approaches are used.

Of importance,  however, is that the  level of agreement  with data in
the  $g$-folding approach  depends on  the quality of what is used for
the underlying model of  structure. Due to the character of the hadron
force, proton  scattering is  preferentially sensitive to  the neutron
matter  distributions  of  nuclei:  a  sensitivity seen  in  a  recent
assessment,    using   proton    elastic   scattering,    of   diverse
Skyrme-Hartree-Fock model  structures for $^{208}$Pb \cite{Ka02}. Such
is the  breadth of success with  target mass and  varied energies that
the  effective $NN$  interactions  have been  established, and so  new
predictions of nucleon  scattering can be considered as  a test of the
quality of the assumed nuclear structure.

Note that  $g$-folding theory of  $NA$ elastic scattering (and  in the
DWA,  of  inelastic  and  charge exchange  reactions)  involves  three
preset quantities. They are
\begin{itemize}
\item An effective $NN$ interaction,
\item The  one  body  density  matrix elements (target structure information), and
\item The  target single-nucleon bound state wave functions.
\end{itemize}
Once specified, a  single calculation of scattering suffices to
obtain a  prediction.  The  effective interaction for 71~MeV  incident
protons has been formed  as is specified  in  the  review~\cite{Am00}. 
Nonetheless a brief review is given next to  allow  interpretation  of 
results of the analyses of $^{6,8}$He ions scattering  from  Hydrogen.
Details of the structures used  are  presented in  the 
following section,        and the results of analyses of the $71A$ MeV
data~\cite{Ha05} are given thereafter.

\section{The effective interaction}

A realistic  microscopic model  of $NA$ reactions  is taken to  be one
that  is  based  upon  $NN$  interactions whose  on--shell  values  of
$t$-matrices (solutions of Lippmann-Schwinger equations) are consistent 
with measured $NN$ scattering data up to the chosen limit energy of 300
MeV. Below pion threshold, the phenomenology of the $NN$ interaction is  
relatively simple, and several one-boson-exchange potential models exist 
with which very good fits have been found to $NN$ phase-shift data.

But  the   presence  of  the  nuclear  medium   alters  such  pairwise
interactions  between an  incoming (positive  energy) nucleon  and any
(bound)  nucleon in the target from the free $NN$ scattering case. The 
result is an effective $NN$ interaction. For incident  proton energies 
to 300 MeV,  it is useful to choose that effective interaction to be a 
mix of central, two-body spin-orbit, and tensor forces, each  having a  
form factor that is a sum of Yukawa functions. Each of those functions  
has a complex, energy, and density dependent strength. The ranges  and 
strengths are obtained by  accurately   mapping  the    double  Bessel 
transform of the effective  interaction  to  the  ($NN$)  $g$-matrices
of the Bonn-B potential. Those $g$-matrices are  the solutions  of the
BBG equations, in momentum space, for energy $E \to k^2$ and for diverse Fermi momenta, $k_F$,
\begin{multline}
g^{JST}_{LL'}\left( p, p'; k, K, k_F \right) = V^{JST}_{LL'}\left( p, p' \right)
+ \frac{2}{\pi} \sum_l \int^{\alpha}_0 V^{JST}_{Ll}\left( p', q \right) \\
\times \left[ \frac{\bar{Q}\left( q, K, k_F \right)}{\bar{E}( k, K, k_F )
- \bar{E}(q, K, k_F ) + i\varepsilon} \right] g^{JST}_{lL'}\left(
q,p;k, K, k_F \right) \; q^2 dq,
\end{multline}
in which $\bar{Q}(q,K,k_F)$ is  an angle-averaged Pauli operator and
$\bar E$  are single  particle energies, all  evaluated at  an average
center-of-mass momentum, $K$. The energy and density dependence of the
complex  effective $NN$ interactions  so formed  have been  crucial in
forming the optical potentials that yield good predictions to measured
data at many energies and from many target nuclei~\cite{Am00}.

\begin{figure}
\scalebox{0.8}{\includegraphics*{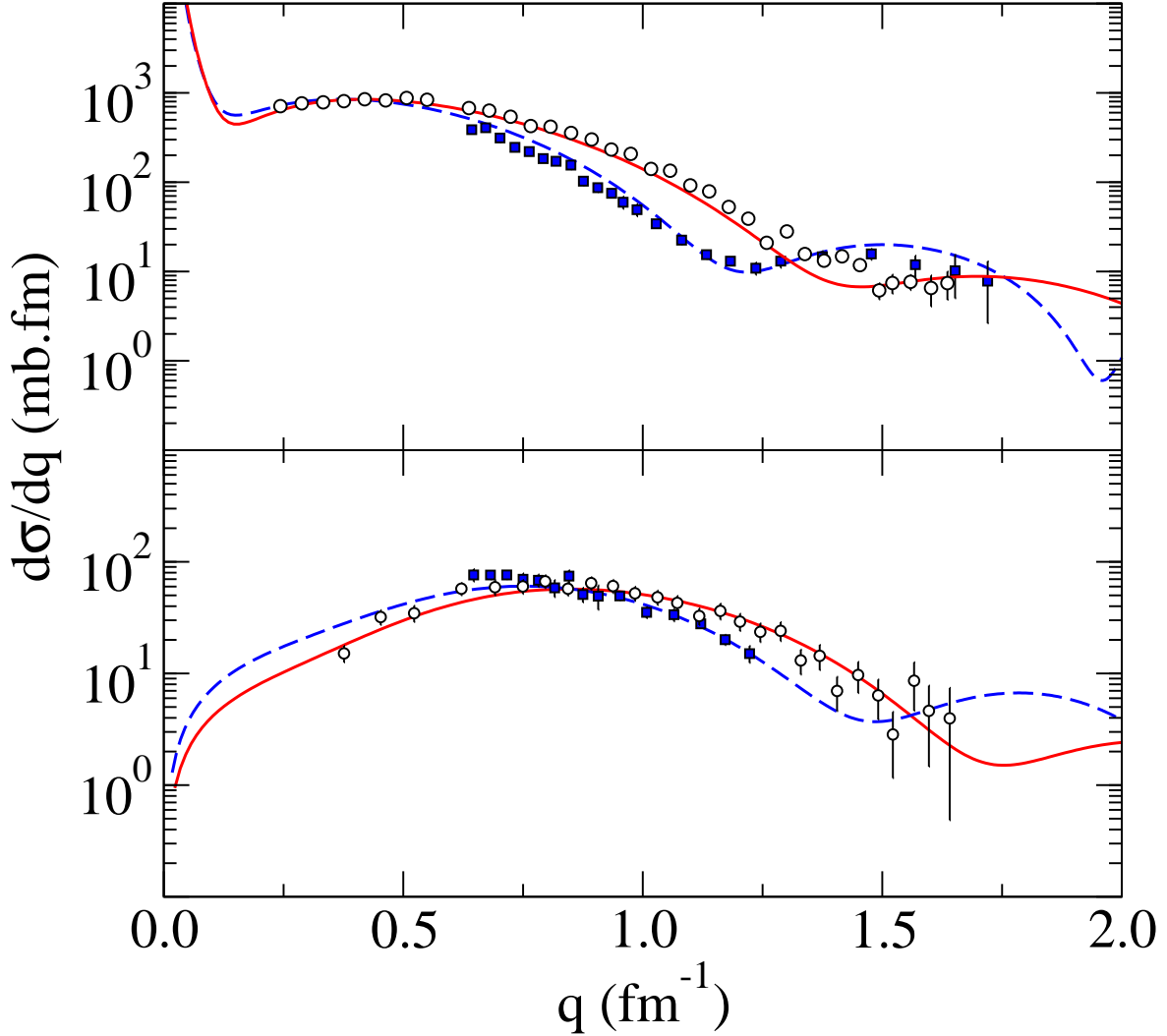}}
\caption{\label{He6-cs-q} (Color online.) Elastic (top) and inelastic (bottom) cross sections of 
$^6$He scattering  from Hydrogen. Data and curves  are identified in
the text.}
\end{figure}

To  illustrate the propriety of this form of effective $NN$ interaction
as well as of the credibility of the $g$-folding approach,       cross
sections for the scattering of  $^6$He ions from Hydrogen are shown as 
functions of momentum transfer in Fig.~\ref{He6-cs-q}. The $g$-folding 
(and DWA for the inelastic scattering) results are compared  with data 
taken at $40.9A$~MeV~\cite{La01} (displayed by the open circles) and 
at $24.5A$~MeV~\cite{St02}  (displayed by the filled squares).
Elastic scattering  data are  shown in  the  top segment and inelastic
scattering data from the excitation of the $2^+_1$ (1.8~MeV) state are
shown in the  bottom segment. The solid and dashed  curves are the
results of predictions of the $40.9A$~MeV and  of the $24.5A$~MeV data 
respectively, made assuming that $^6$He has a  neutron halo.       The  
inelastic scattering results were obtained using the (predictive)  DWA 
approach   in   which  the  distorted  waves  were  generated from the 
$g$-folding optical potentials with which the elastic scattering cross
sections were well-reproduced and the transition OBDME were obtained from the 
no-core shell model calculation that also gave the ground state OBDME.
The same Woods-Saxon (WS) bound state functions that gave  a   neutron  
halo and lead to the good cross sections for elastic scattering   were
also used  as  the  single-particle (SP) wave  functions for  both the  
ground and excited  state. The effective $NN$ interaction used to form
the $g$-folding optical potentials was also used   as   the transition 
operator effecting the inelastic scattering. Consequently all details
describing the inelastic excitation within the DWA have been preset so
that inelastic scattering magnitudes, in particular, are  predictions. 
No core polarization, or other scaling values,    are needed to obtain 
agreement  with data.  In  fact, it  is  the magnitude  of the  forward 
scattering angle (low momentum  transfer) inelastic cross section data
that is prime evidence of the neutron  halo  properties  of the $^6$He
states~\cite{La01,St02}. That the overall (shell model) approach is valid
is confirmed by the self-consistent and predictive descriptions of the elastic, inelastic,
and reaction cross sections \cite{La01}.  The  transition  form factor  for  inelastic
scattering  stresses  the  properties  of  the  nuclear  surface  and,
consistent with  that, extension of the neutron  matter influences low
momentum  transfer aspects  of that  scattering. In  contrast,  and as
shown by  the lower energy results~\cite{La01,St02},  the neutron halo
effects in the  elastic scattering cross sections are  most evident at
momentum  transfer  values   greater  than  $\sim  1$~fm$^{-1}$.  This
reflects the  influence of  the bulk volume  properties of  the $^6$He
wave functions  in defining the  optical potentials: it  is the
depletion of  neutron density  in the inner  region, coming from the extension of
the neutron density at the surface to large radii, that results in the reduction of the differential
cross section at large momentum transfers.

As the same structure (OBDME, SP wave functions, etc.) was used in the
calculations at  both energies, the results plotted  here as functions
of momentum  transfer not only  confirm the neutron-halo  character of
$^6$He but also reveal that the effective interaction varies radically
with energy and  density in a way well-defined by  the BBG approach to
its definition.   Thus predictions for the scattering at $71A$~MeV can 
be expected to have high probability  of matching  data.   Indeed  the
predicted  cross  section  does  match  existing         cross-section 
data~\cite{Ko97} as will be shown again later.       The new data, for 
analyzing powers~\cite{Ha05} are considered now since those are   very 
sensitive to details of the scattering matrices.      One might expect 
predictions of the analyzing power at $71A$~MeV to be realistic  since 
very good agreement has been found with $g$-folding  model predictions  
spin observables with data from the scattering  of 65 and 200~MeV protons from 
many stable nuclei~\cite{Am00}.     However, spin observables are very 
sensitive to details in scattering theory,    with analyzing powers in 
particular being so on specifics of structure~\cite{Am00}.

\section{Structure of $^{6,8}$He}

$^6$He  and $^8$He are  among the most studied  of the exotic  systems 
given  that they  lie close  to both  the valley of stability and  the 
neutron drip line.      They are Borromean since $^{5,7}$He are both 
neutron-unstable.    $^6$He is classed as  a two-neutron halo nucleus,
though the term is perhaps unfortunate: rather, it should be viewed as a nucleus
with an exceptionally extended neutron density.   $^8$He is not a 
neutron-halo nucleus: it is a neutron-skin nucleus. Both descriptions are
supported by analyses of existing proton scattering data \cite{Ka00,La01}.

For the purposes of the present study, a complete $(0+2+4)\hbar\omega$ 
shell model  has  been  used \cite{Ka00} to describe both $^6$He and $^8$He. 
The Zheng~\cite{Zh95} $G$-matrix shell model interaction,   as
derived  from  the  Nijmegen  III $NN$  interaction~\cite{St94},   was 
used to define the Hamiltonians.  The structures  so defined  not only
provide the OBDME (or occupation numbers) defining the  density in the
case   of  elastic  scattering and  transition OBDME    for  inelastic
scattering  but also  the  relevant SP  wave  functions.    The Zheng 
$G$-matrix  interaction  requires a  specification  of the  oscillator
energy to obtain the two-body potential energy matrix elements so that
energy specifies  the oscillator parameter  for use in  the scattering
calculations.    We have used   $\hbar\omega    =   14$~MeV   in   the
$(0+2+4)\hbar\omega$ space calculations. But all such shell model wave
functions do  not give a neutron-halo  character. At best,  and due to
the  Gaussian SP  functions involved,  the density  profiles  from the
shell model information form neutron (proton) skins.

In the case of $^6$He, established now as a halo nucleus~\cite{La01},
oscillators need be replaced with, for example,  WS wave functions  to
effect an extended neutron matter character. (This was necessary to reproduce
the $B(E1)$ value in $^{11}$Be \cite{Mi83}.)  By choosing the binding 
energy of  the halo-neutron orbits to be fixed by the  single  neutron 
separation energy to the lowest energy resonance in $^5$He (1.8~MeV), 
the  associated density profile has the extensive neutron density that
is characterized as a halo.  To date that has sufficed  to  replicate 
cross section data  for  many  incident  ion  energies,  even at 
$15A$~\cite{Ka07} with $^8$He, where the use of WS functions describes well
the skin of $^8$He.

\section{Differential cross sections and analyzing powers}

$g$-folding model results (of elastic scattering)  are pertinent for
calculated differential   cross  sections  exceeding a few tenths of a 
mb/sr.          Typically that range for 71~MeV protons reaches 100 to 
110$^{\circ}$ in the center of mass.

\begin{figure}
\scalebox{0.8}{\includegraphics*{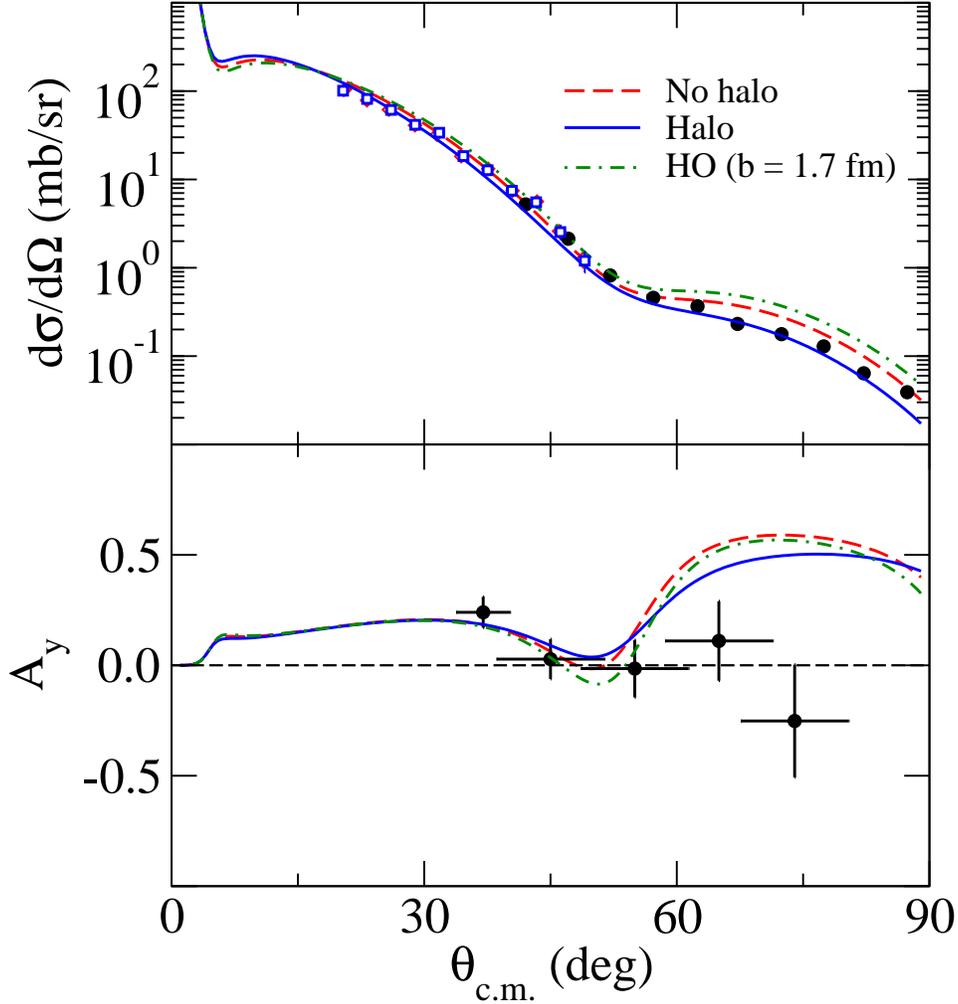}}
\caption{\label{He6-71} (Color online) Differential  cross sections  (top) and  analyzing powers
(bottom) for the scattering of $71A$~MeV $^6$He ions from hydrogen.}
\end{figure}

The differential cross sections and  analyzing powers for the elastic scattering of 
$^6$He ions from Hydrogen are shown in Fig.~\ref{He6-71}. 
The results of three calculations are presented, the difference is in the
choice of single particle wave functions used. The solid line is the result
using the WS set as in Fig.~\ref{He6-cs-q}, while the dashed and dot-dashed
lines show the results of the calculations made using WS functions with a larger
binding energy ($\sim 7$~MeV) for the neutrons in the $0p$-shell (hereafter known
as the ``nohalo'' result) and also with oscillator functions ($b = 1.7$~fm),
respectively. The data are from Sakaguchi \textit{et al.}, \cite{Sa11}, and
Korshenninkov, \textit{et al.} \cite{Ko97}, portrayed by the circles and squares,
respectively. In the case of the differential cross section, the data 
to $80^{\circ}$ are not sensitive enough to distinguish between the three
results, although there may be a slight preference for the halo result
at larger angles. This is in contrast to the analyses of proton scattering at
lower energies which definitively show the halo \cite{La01,St02}.

For the analysing power of $^6$He, all calculations reproduce the data at forward angles.
At larger angles, all models predict a positive analysing power, while the data
indicate an analysing power that is flat and near zero. However, it is
noteworthy that the data were obtained by binning over $10^{\circ}$ in angle
to obtain each datum \cite{Sa11}, as shown in the figure. That binning may give rise to a problem in
establishing the analysing power at these angles: the averaging inherent in the binning,
especially over such a large angular range may yield a smaller value in the analysing
power. This is especially critical as the analysing power is a sensitive cancellation of
terms in the scattering amplitude, normalised to the differential cross section. There is also an 
additional uncertainty of 19\% in the scale of the measured
analysing power \cite{Sa11}. As such, further experiments may be necessary to provide better data 
by which to evaluate the analysing power.

\begin{figure}
\scalebox{0.8}{\includegraphics*{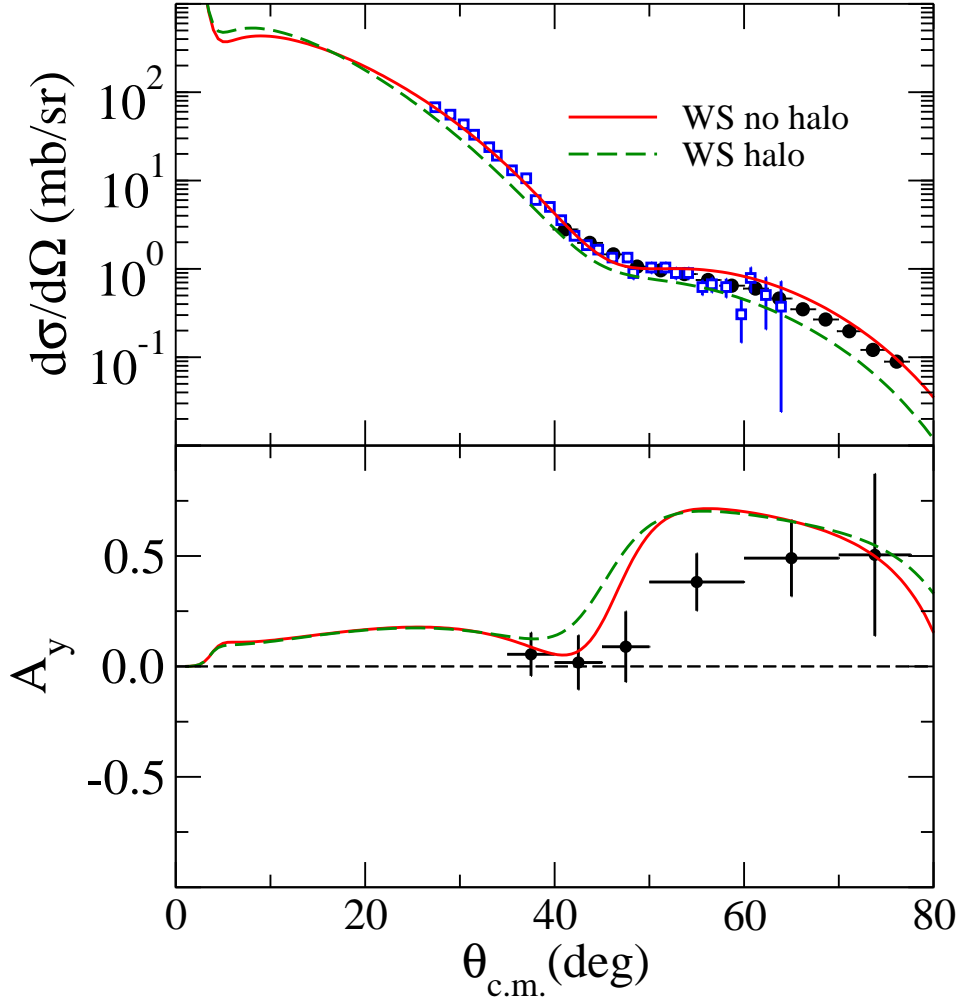}}
\caption{\label{He8-71} (Color online) Differential  cross sections and  analyzing powers  for the
scattering of $71A$~MeV $^8$He ions from Hydrogen.}
\end{figure}

The results for elastic scattering of $^8$He ions from Hydrogen are 
presented in Fig.~\ref{He8-71}, where the data portrayed are from Sakaguchi \textit{et al.}
\cite{Sa13} (circles) and Korshennikov \textit{et al.} \cite{Ko97} (squares).
In this case, the WS calculation uses the pertinent single particle wave functions
with the correct single-nucleon binding energies for $^8$He \cite{Am00}, the result of which
is displayed by the solid line. In this case, the result is termed
``no halo'', and that result agrees well with the data. The ``halo'' result corresponds, somewhat
arbitrarily, to the setting of those single-nucleon binding energies much lower, more in
line with that of $^6$He, as we had done previously \cite{Am00}. The data show
a preference for the no halo result, consistent with the description of $^8$He nucleus
as that having a neutron skin.

Unlike the analysis of the cross section data, the analysis of the analysing power data
show little difference between the two results, save for there being a preference
again for the no-halo case. The data still have the problem of binning,
as was the case for scattering from $^6$He. There is better agreement with the data in this
case. More data may be needed for both scatterings, however.

\section{Conclusions}

We have presented analyses of analysing power data for the scattering of intermediate energy $^6$He
and $^8$He ions from hydrogen. The $g$-folding approach, based on the use of multi-$\hbar\omega$
no-core shell model wave functions, with appropriate WS single-particle wave functions, gives
agreement with both differential cross-section and analysing power data. Central to this
level of agreement is the constraint of the binding energy of the WS functions to the
separation energy of the single nucleon from the target. This applies to both $^6$He and $^8$He,
the former giving rise to an extensive neutron density, otherwise called the halo. The neutron
skin in $^8$He is also reproduced.

Given that the analysing powers are very sensitive to the details of the interaction between the
projectile and target, it is tempting to interpret the data as indicating new physics in
describing that interaction. However, while the general shapes of the analysing powers are
reproduced by the calculations from the microscopic model, more data will be needed to
make conclusive statements.

We would like to thank S. Sakaguchi for useful discussions. SK acknowledges support of the 
National Research Foundation of South Africa.

\bibliography{He68-Ay-paper}

\end{document}